\documentclass[12pt,preprint]{aastex}

\slugcomment{Accepted for publication in ApJ}

\shorttitle{Interstellar Meteors}
\shortauthors{Musci et al.}

\begin{document}


\title{An Optical Survey for mm-Sized Interstellar Meteoroids}


\author{R. Musci, R.~J. Weryk, P. Brown, M.~D. Campbell-Brown, and P.~A. Wiegert}
\affil{Department of Physics and Astronomy, University of Western Ontario,
    London, Ontario N6A 3K7, Canada}
\email{rmusci@uwo.ca}


\begin{abstract}
We report high resolution multi-station observations of meteors by the Canadian Automated Meteor Observatory (CAMO) recorded from June 2009 to August 2010. Our survey has a limiting detection magnitude of +5\,mag in $R$-band, equivalent to a limiting meteoroid mass of $\sim2\times$10$^{-7}$\,kg. The high metric trajectory accuracy (of the order of 30\,m perpendicular to the solution and 200\,m along-track) allows us to determine velocities with average uncertainty of $<$\,1.5\% in speed and $\sim$0.4$\degr$ in radiant direction. A total of 1739 meteors had measured orbits. The data has been searched for meteors in hyperbolic orbits, which are potentially of interstellar origin. We found 22 potential hyperbolic meteors among our sample, with only two of them having a speed at least three sigma above the hyperbolic limit. For our one year survey we find no clear evidence of interstellar meteoroids at mm-sizes in a weighted time-area product of $\sim$10$^4$\,km$^2$\,h. Backward integrations performed for these 22 potentially hyperbolic meteors to check for close encounters with planets show no considerable changes in their orbits. Detailed examination leads us to conclude that our few identified events are most likely the result of measurement error. We find an upper limit of $f_{\rm ISP}<2\times$10$^{-4}$\,km$^{-2}$\,h$^{-1}$ for the flux of interstellar meteoroids at Earth with a limiting mass of $m>2\times10^{-7}$\,kg.
\end{abstract}


\section{Introduction}

Direct measurements of interstellar particles (ISPs) in our Solar System are of significant astrophysical importance. Since direct measurements of ISP spatial densities are very difficult \citep{tes2003}, atmospheric meteor detection is perhaps the only technique that allows direct flux measurements for ISPs at large ($>$\,10$^{-8}$\,kg) sizes. ISPs at large sizes maintain ``memory'' of their originating sources, being negligibly perturbed in the ISM by other forces; it becomes possible to link larger ($>$10$\mu$m) ISPs to specific sources, providing a bridge between astrophysical studies of circumstellar dust and in situ measurements \citep[e.g.,][]{mur2004}.

The value of the detection of ISPs as meteors has been understood for almost a century. The entire concept of ISPs and meteors was the motivation for many of the earliest instrumental meteor observations \citep{hug1982}. Detection of ISPs, however, hinges on proper error estimates, values seldom fully explored for most instruments used for meteor observations.

The discussion about the influx of ISPs at Earth started with the publication of a visual meteor catalog by \citet{v&h1925}. The authors found hyperbolic orbits, i.e., heliocentric speeds $v_{\rm h} \gtrsim 42.1$\,km\,s$^{-1}$, for 79\% of observed meteors. Whereas this result was confirmed by some authors, e.g., \citet{opi1950}, others claimed that there was no evidence for hyperbolic orbits due to large measurement errors \citep{por1943, por1944, whi1954}. \citet{j&w1961} found no clear indication of hyperbolic meteors for the most precise orbits measured by the Super-Schmidt cameras in the 1950's - 1960's. The expected heliocentric speed for ISPs at Earth depends on the stellar velocity of the originating star system relative to the Sun, a value typically $\sim$20\,km\,s$^{-1}$, implying $v_{\rm h,ISP} > 46$\,km\,s$^{-1}$ at Earth \citep{opi1950}. This does not mean that ISPs with lower velocities cannot exist, but that they are not expected to be common from purely dynamical considerations and would be difficult to detect without very precise velocity determination, being close to the hyperbolic limit and likely confused with the large population of nearly unbound cometary meteoroids. Note that while $v_{\rm h,ISP}$ is expected to be $>$46\,km\,s$^{-1}$, an interstellar source of meteoroids can produce in-atmosphere velocities as low as 15\,km\,s$^{-1}$. In general, true ISPs are not expected to have unusually high velocities in Earth's atmosphere due to the random collision geometry relative to Earth's orbital motion.

The existence of very small ISPs in the solar system was confirmed by the Galileo and Ulysses probes \citep{lan2000}. For larger masses ($m>10^{-12}$\,kg), however, the true flux of ISPs remains unclear, with \citet{haj2008} suggesting that many published hyperbolic orbits may be the consequence of measurement errors, while also noting that hyperbolic meteors could be produced by planetary perturbations. \citet{j&s1985} suggested that planetary perturbations or collisions may change bound meteoroid orbits into hyperbolic ones, though no clearly unbound orbits with recent close planetary encounters have been identified to date.

Here we report on a one year survey for ISPs using a two station automated electro-optical meteor observatory, providing a large dataset of meteors with high metric accuracy compared to other video systems ($\sim$0.4$\degr$ radiant error, $\sim$1.5\% atmospheric speed error) and masses of the order of $10^{-7}$\,kg. We discuss possible hyperbolic candidates identified during the survey's $\sim$10$^4$\,km$^2$\,h collecting area-time product. The orbits of each hyperbolic candidate are analyzed to check for prior close encounters with the major planets. Finally, we present an estimate of the flux of ISPs with $m>2\times10^{-7}$\,kg at Earth obtained from our survey.

\section{Previous Studies}

The first unambiguous detections of ISPs in the Solar System were from the dust experiment on the Ulysses spacecraft \citep{gru1993}. They found micrometer-sized grains moving with high velocities and appearing to emanate from the direction of the local interstellar gas flow, a result later confirmed by the Galileo mission \citep{bagu1995}. \citet{bagg1993} reported the first radar detection of micron-sized hyperbolic meteors with the Advanced Meteor Orbit Radar (AMOR). They noted a well distributed hyperbolic background influx and a discrete stream of ISPs they attributed to the dust debris-disk star $\beta$ Pic \citep{bagg2000}. \citet{mat1999} reported detection of an interstellar meteor with the Arecibo Observatory radar. They later claimed to have identified 143 ISPs \citep{mei2002a}. Their results, however, remain controversial. First, they had assumed that all meteors came down the main beam and were not in one of the sidelobes. Second, meteors crossing the main beam come in at an angle that cannot be measured, meaning that only a radial velocity is truly known. Both cases lead to large uncertainties in velocity. The data gathered by the Canadian Meteor Orbit Radar (CMOR) was analyzed for ISPs with $m>10^{-8}$\,kg \citep{wer2004}. Out of 1.5 million measured orbits, they found 12 possible events when (the large) measurement errors were taken into account. \citet{haj1994} searched the photographic database of the IAU Meteor Data Center for evidence of ISPs. She concluded that the vast majority of the apparent hyperbolic meteors were a consequence of measurement error. The most precise catalogues may include some true hyperbolic meteors, however, the hyperbolic excess $\Delta v_{\rm h}$ of the speed, i.e., the amount above the hyperbolic limit, was smaller than expected based on average relative stellar velocities. Several updates of that work have led to the same conclusion \citep[e.g.,][]{haj2002, haj2008}. An analysis of the Japanese meteor shower catalogue from video observations \citep{haj2011} also showed no evidence for interstellar meteoroids. Two ISPs detected with image-intensified video cameras were reported by \citet{h&w1997}. The measured $v_{\rm h}$ of those events are several times the error values above the hyperbolic limit and also 2$\sigma$ and 3$\sigma$ above the expected $v_{\rm h, ISP}$.

\begin{figure}
\plotone{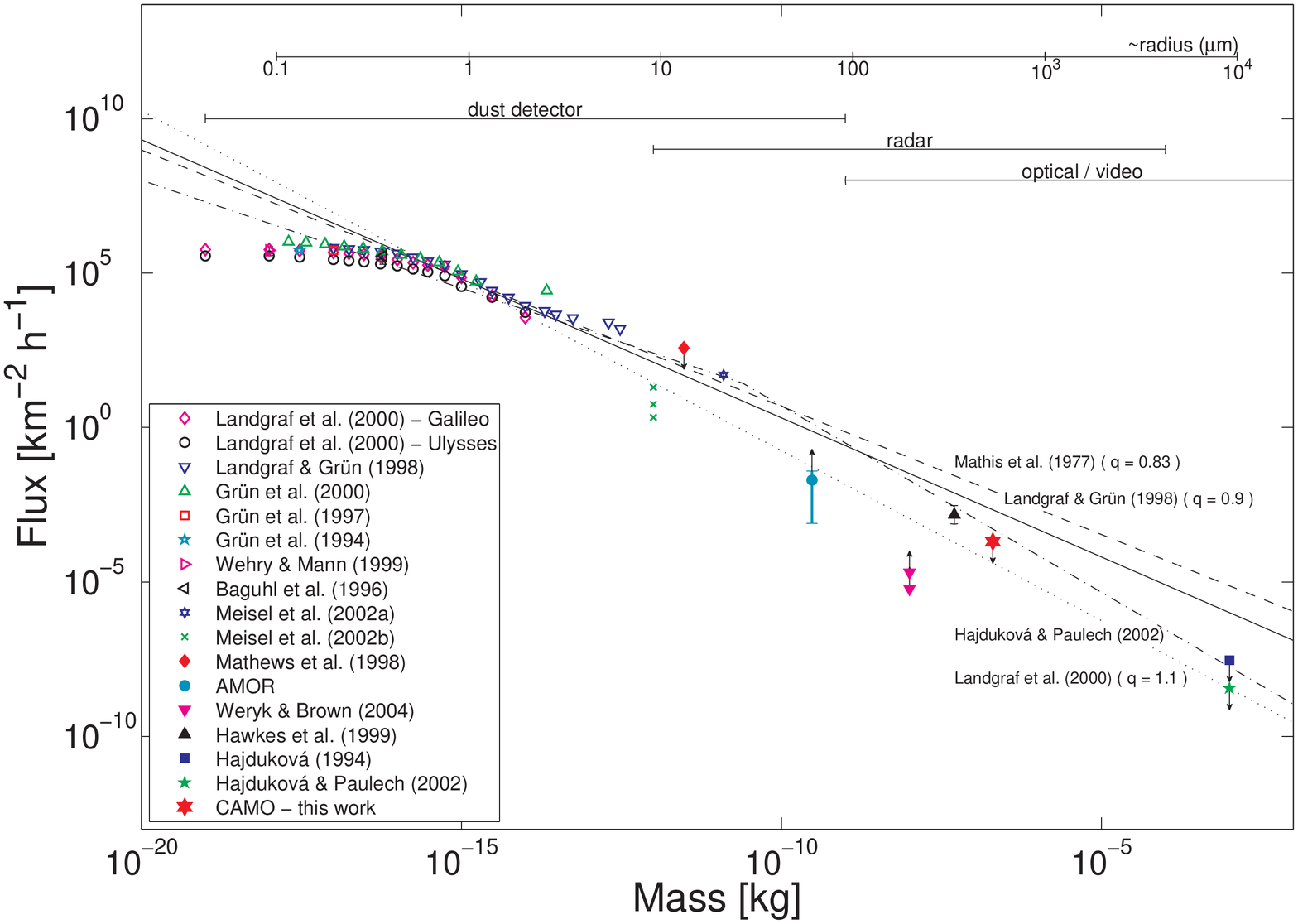}
\caption{Interstellar meteoroid flux estimates from various studies. The large star represents our result. Arrows indicate upper and lower limits. The data points from \citet{mei2002b} are for Geminga supernova particles assuming different models and fits. The AMOR data includes the results from \citet{bagg1993}, as well as the interpretation of those data from \citet{tay1996} and \citet{bagg2000}. The points from \citet{wer2004} are for $v_{\rm h}>2\sigma$ and $v_{\rm h}>3\sigma$, respectively, above the hyperbolic limit. The ranges at the top of the figure give the approximate sensitivity for different detectors. For comparison, the lines with different styles represent the mass distribution from several models and power-law fits. The slope from \citep{mrn1977} is identical with the collisional cascade model \citep{doh1969, tan1996, wya2007}.\label{fig1}}
\end{figure}

The flux, i.e., the number of particles observed in a given area per unit time, for ISPs reported by various studies was calculated and compared by \citet{haj2002} and updated in \citet{haj2006}. We took those results as the basis for an overview of ISPs fluxes at Earth reported in the literature in the mass range from $10^{-20}$\,kg to $10^{-2}$\,kg (Figure~\ref{fig1}\notetoeditor{Figure 1 should be full page width in print}). We calculated the fluxes based on information from primary sources, including results from dust detectors on spacecraft, as well as radar and optical measurements of meteors. These are shown in Figure~\ref{fig1}, including several published power-law models and fits of the form $N(m)\propto m^{-q}$. The fit from \citet{haj2002}, which is based on the interplanetary flux models from \citet{fech1973} and \citet{div1993}, includes a break at about $2.4\times10^{-11}$\,kg. From Figure 4 in \citet{haj2002} we estimate $q\approx0.7$ for smaller and $q\approx1.2$ for larger particles. It has to be noted that the later values are empirical and not strongly supported by modeling.

\section{Data Collection System \& Reduction Methodology}
The cameras used for our survey are part of the Canadian Automated Meteor Observatory (CAMO). The CAMO consists of identical camera systems located at two sites separated by 45\,km, the Elginfield Observatory (43.1928$\degr$N, 81.3157$\degr$W) and a site near Hickson, ON (43.2642$\degr$N, 80.7721$\degr$W), mounted inside environmentally controlled housings. The two camera systems are pointed at a fixed azimuth and altitude at each site, observing a common volume of atmosphere near 100 km altitude. This geometry allows determination of a trajectory and therefore an orbit for each meteor observed from both sites.

The systems have image intensified video cameras, with an ITT Nitecam model 380 generation 3 image intensifier coupled to a PCO Imaging PCO.1600 CCD video camera. The camera has 1600$\times$1200\,pixels, though only 1024$\times$1024 are currently used due to the optical configuration. A 50\,mm f/0.95 Navitar lens is attached to the intensifier, giving a field of view (FOV) of $\sim$21\degr. The pixel scale is 74\arcsec\,pix$^{-1}$ corresponding to $\sim$35\,m at an altitude of 100\,km. The cameras run at 20 frames per second and have a bit depth of 14-bits. The limiting $R$-band meteor detection peak magnitude of the system is +5\,mag. We used the color index $(V-R)=0.72$ from \citet{kik2011} to convert the $R$ magnitudes to $V$ magnitudes and a brightness dependent color index $1<(V-pg)<2$ from \citet{jac1957} to convert the $V$ magnitudes to photographic magnitudes. Three different methods to calculate the luminous efficiency $\tau$ were used to estimate the limiting meteoroid mass: a constant $\tau$ of $0.7\%$, and a variable $\tau$ depending on the speed of the meteor based on \citet{hil2005} and \citet{c&m1976}. The first two $\tau$ are for $V$ magnitudes, whereas the latter is for photographic magnitudes. The resulting limiting meteoroid mass is $\sim2\times$10$^{-7}$\,kg. The metric trajectory accuracy of the system is $\sim$30\,m perpendicular to the mathematical solution of the trajectory. This value was determined from 16 meteors that were observed with an additional camera at a third site. The accuracy of our system is comparable to the one of the Baker Super-Schmidt cameras of the Harvard Meteor Project \citep{m&p1961}. It is, however, slightly less accurate than the photographic systems used within fireball networks \citep[e.g.,][]{kot2006}, which can reach an accuracy of less than 15 meters.

Whenever the conditions permit, the system automatically acquires data. All images are searched for meteors using the MeteorScan program \citep{gur2008}, which are then stored for additional manual processing. To determine the efficiency of the MeteorScan program, several hours of raw video data were searched for meteors by an analyst and compared with the results from MeteorScan. Although MeteorScan misses the faintest meteors, we found that $\sim$95$\%$ of the meteors brighter than +5\,mag are successfully found.

\section{Analysis} \label{s_ana}

Events detected at both sites were processed by manually choosing the apparent meteor trail head on each frame. The plate fits for the picks were based on at least 20 randomly selected and scattered stars within the FOV. The trajectories were determined using the non-linear trajectory fit model described in \citet{bor1990}. The heliocentric velocities and the orbits of the meteoroids were computed following \citet{cep1987}.

During the analysis, we determined that the results strongly depend on the subjective ``pick'' of the fiducial point that best represents the position of the meteor. We had several experienced analysts (up to four) process the same events in order to compare the solutions and to confirm the robustness of the resulting hyperbolic orbits. For all potential hyperbolic events that had been processed by four analysts at least one of the solutions was not hyperbolic. One example is shown in Table~\ref{tab1}. The analysis of multiply processed events showed that the average difference in the picks by different analysts is 1-2 pixels. Considering the pixel scale of the system, this is $<$\,100\,m. These findings support the notion that other marginal detections of hyperbolic meteors with optical instruments are strongly influenced by such small subjective measurement errors, a conclusion also reached by \citet{whi1954}.

\begin{deluxetable}{lrrrrrrrrr}
\tabletypesize{\scriptsize}
\tablecaption{An example of a set of solutions for an event processed by different analysts\label{tab1}}
\tablewidth{0pt}
\tablehead{
\colhead{\#} & \colhead{RMS} & \colhead{$h_{\rm B}$} & \colhead{$h_{\rm E}$} & \colhead{$Q$} &
\colhead{$v_0$} & \colhead{$v_{\rm hwp}$} & \colhead{$v_{\rm g}$} & \colhead{$v_{\rm h}$} & \colhead{$\Delta v_{\rm h}$} \\
\colhead{} & \colhead{(m)} & \colhead{(km)} & \colhead{(km)} & \colhead{(deg)} &
\colhead{($\rm km\,s^{-1}$)} & \colhead{($\rm km\,s^{-1}$)} & \colhead{($\rm km\,s^{-1}$)} & \colhead{($\rm km\,s^{-1}$)} & \colhead{($\rm km\,s^{-1}$)}
}
\startdata
 1 & 24.9 & 126.47$\pm$0.09 & 94.88$\pm$0.09 & 7.82 & 41.70$\pm$0.18 & 42.60$\pm$1.45 & 40.97$\pm$1.50 & 43.90$\pm$1.11 & 1.43 \\
 2 & 14.0 & 121.39$\pm$0.06 & 95.09$\pm$0.06 & 7.64 & 39.95$\pm$0.23 & 40.08$\pm$0.46 & 38.35$\pm$0.48 & 41.86$\pm$0.36 & $-$0.62 \\
 3 & 18.0 & 122.06$\pm$0.08 & 94.99$\pm$0.07 & 7.65 & 41.07$\pm$0.09 & 42.43$\pm$1.10 & 39.75$\pm$1.14 & 42.96$\pm$0.84 & 0.48 \\
 4 & 13.7 & 125.38$\pm$0.06 & 95.15$\pm$0.05 & 7.73 & 40.41$\pm$0.09 & 41.16$\pm$0.18 & 39.47$\pm$0.18 & 42.69$\pm$0.16 & 0.21 \\
\enddata
\tablecomments{RMS is the root of the mean squared residuals of the trajectory solution (a proxy measure for the internal consistency of the measured points), $h_{\rm B}$ and $h_{\rm E}$ are begin and end height, $Q$ is the convergence angle, $v_0$ is the average measured velocity, $v_{\rm hwp}$ is the average velocity half-way along the observed trail, $v_{\rm g}$ is the geocentric velocity, $v_{\rm h}$ is the heliocentric velocity, and $\Delta v_{\rm h}$ is the hyperbolic excess.}
\end{deluxetable}

To establish error bounds on each event, we used a Monte Carlo method to analyze the error in the trajectory and velocity calculation. We took all fiducial pixel locations and applied Gaussian distributions with a standard deviation of 1 pixel both randomly and systematically. These new locations were then run through the trajectory solver. The resulting distributions of radiant positions and speed were fit with Gaussian profiles and the widths of these distributions were used as our uncertainty $\sigma$. The resulting average $\sigma$ for the velocity is $\overline{\sigma}\approx1.5\%$ for the events detected during the survey presented in this study. This is in good agreement with the average along track uncertainty of 200\,m, which is based on the trajectory solutions of the same events. 

The system collecting area was numerically computed for each possible radiant direction after correcting for meteor range, camera sensitivity, multi-station geometry, and meteor detection sensitivity. Typical collecting areas per radiant direction are $\sim$30\,km$^2$. The procedure and its validation against known sporadic meteor flux is described elsewhere \citep{cam2011}.

\section{Results} \label{s_res}

The flux of ISPs is defined as
\begin{equation}
f_{\rm ISP} = \frac{N_{\rm ISP}}{t\cdot A_{\rm col}},
\end{equation}
where $N_{\rm ISP}$ is the number of observed ISPs, $t$ the total observation time, and $A_{\rm col}$ the collecting area. Between 2009 June 4 and 2010 August 17, the systems from both sites were observing simultaneously for more than 300 hours. For radiants visible to our system we obtain an average $A_{\rm col}$ per radiant of $30\pm14$\,km$^2$ and a corresponding time-area product of $9000\pm4200$\,km$^2$\,h.

A total of 1739 meteors had measured velocities, most observed in late Summer and early Spring. Only 143 meteors were observed from October to February, mainly due to poor weather conditions. Hyperbolic orbits were found for 115 meteors during the first round of processing. Many of these solutions were unreliable and rejected for various reasons. For our data, we found that the solutions were not well defined if the convergence angle $Q$ between the station intersecting planes was less than $5\degr$. Poor solutions also resulted if the number of measurable frames (picks) from either site was $<$\,5. As a final quality control, we required the difference between the average velocities of the trajectory from the two sites to be $|(v_1-v_2)/v_1|<0.02$. After applying these quality control factors, only 22 potential hyperbolic meteors remained. Some measured parameters for these 22 meteors are given in Table~\ref{tab2}. The hyperbolic limit for meteoroids at Earth varies between $41.78$\,km\,s$^{-1}$ and $42.49$\,km\,s$^{-1}$, depending on the position of Earth on its orbit. Of our final 22 potential hyperbolic meteors, only two had $v_{\rm h}>3\sigma$ above the limit and six had $2\sigma<v_{\rm h}<3\sigma$, with the largest value being $\sim$13$\sigma$. Three of the other candidates could be associated with a meteor shower (one with the November Omega Orionids and two with the Perseids).

\begin{deluxetable}{rrrrrrrrrrrrrrrr}
\tabletypesize{\scriptsize}
\rotate
\tablecaption{Potential hyperbolic meteors\label{tab2}}
\tablewidth{0pt}
\tablehead{
\colhead{$\lambda_{\rm S}$} & \colhead{$\alpha_{\rm R}$} & \colhead{$\delta_{\rm R}$} & \colhead{$h_{\rm B}$} & \colhead{$h_{\rm E}$} &
\colhead{$Q$} & \colhead{$M_{\rm R}$} & \colhead{$m$} & \colhead{$v_{\rm 0}$} & \colhead{$v_{\rm h}$} & \colhead{$\sigma$} &
\colhead{$q$} & \colhead{$e$} & \colhead{$i$} & \colhead{$\omega$} & \colhead{$\Omega$} \\
\colhead{(deg)} & \colhead{(deg)} & \colhead{(deg)} & 
\colhead{(km)} & \colhead{(km)} & \colhead{(deg)} & \colhead{(mag)} & \colhead{(g)} & \colhead{($\rm km\,s^{-1}$)} & 
\colhead{($\rm km\,s^{-1}$)} & \colhead{} & \colhead{(AU)} & \colhead{} & \colhead{(deg)} & \colhead{(deg)} & \colhead{(deg)}
}
\startdata
118.8 &  23.98 &  -5.98 & 118.35 & 105.04 & 16.2 & -1.0\hspace{5 pt} & 8.88$\times10^{-3}$ & 70.03$\pm$0.11 & 43.93$\pm$0.06 & 13.46 & 0.96 & 1.20 & 154.11 &  26.05 & 298.59 \\
145.4 &  10.96 &  57.79 & 118.81 &  90.55 &  6.1 & -0.2\hspace{5 pt} & 6.27$\times10^{-3}$ & 57.59$\pm$1.00 & 44.29$\pm$0.73 &  2.11 & 0.96 & 1.23 & 100.77 & 205.35 & 145.22 \\
152.1 & 331.76 &  37.98 & 103.14 &  91.33 & 33.3 &  2.2\hspace{5 pt} & 1.52$\times10^{-3}$ & 38.33$\pm$0.15 & 42.00$\pm$0.30 &  0.27 & 0.62 & 1.01 &  52.16 & 252.86 & 151.97 \\
152.2 &  64.47 &   3.48 & 117.50 & 106.17 &  8.8 &  0.9\hspace{5 pt} & 2.17$\times10^{-3}$ & 59.22$\pm$0.55 & 42.24$\pm$0.27 &  1.27 & 1.02 & 1.03 & 149.50 & 357.37 & 332.08 \\
152.3 &  52.47 &   1.48 & 118.93 &  99.60 & 19.9 &  0.1\hspace{5 pt} & 2.67$\times10^{-3}$ & 37.19$\pm$2.63 & 43.38$\pm$0.67 &  2.23 & 0.90 & 1.13 & 149.88 &  36.04 & 332.17 \\
177.6 &  70.30 &  10.69 & 135.08 &  98.27 & 20.0 & -2.9\hspace{5 pt} & 4.02$\times10^{-2}$ & 69.74$\pm$0.01 & 42.77$\pm$0.14 &  2.77 & 0.77 & 1.05 & 158.24 &  56.54 & 357.41 \\
177.6 &  33.54 &  62.76 & 108.20 &  94.32 & 70.7 &  4.3\hspace{5 pt} & 1.10$\times10^{-4}$ & 49.55$\pm$0.90 & 44.56$\pm$0.39 &  2.99 & 0.81 & 1.20 &  94.40 & 229.75 & 177.45 \\
181.5 &  65.02 &  12.94 & 105.21 &  86.56 & 24.1 & -2.8\hspace{5 pt} & 4.63$\times10^{-2}$ & 65.95$\pm$0.20 & 42.79$\pm$0.71 &  1.04 & 0.52 & 1.04 & 161.27 &  85.94 &   1.34 \\
182.3 &  65.11 &  41.36 & 116.97 &  95.49 & 12.6 & -0.1\hspace{5 pt} & 3.77$\times10^{-3}$ & 66.22$\pm$0.39 & 43.48$\pm$0.40 &  1.86 & 0.72 & 1.10 & 142.03 & 243.01 & 182.16 \\
182.5 & 133.45 &  37.09 & 114.36 &  93.81 & 11.0 & -0.4\hspace{5 pt} & 4.84$\times10^{-3}$ & 61.34$\pm$0.41 & 42.43$\pm$0.41 &  0.68 & 0.45 & 1.02 & 134.28 &  84.16 & 182.34 \\
228.9 & 168.64 &  62.94 & 110.28 & 100.38 & 15.7 &  2.9\hspace{5 pt} & 8.80$\times10^{-4}$ & 54.33$\pm$0.88 & 42.62$\pm$0.43 &  0.44 & 0.97 & 1.03 &  94.74 & 184.42 & 228.80 \\
229.0 & 146.62 &  78.66 & 109.82 &  94.68 &  7.7 &  2.1\hspace{5 pt} & 2.46$\times10^{-3}$ & 49.79$\pm$0.18 & 43.06$\pm$0.45 &  0.20 & 0.92 & 1.06 &  76.51 & 211.71 & 228.84 \\
247.3 &  92.19 &  16.16 &  99.32 &  82.97 & 18.8 &  1.9\hspace{5 pt} & 1.50$\times10^{-3}$ & 43.41$\pm$0.06 & 42.81$\pm$0.32 &  1.21 & 0.11 & 1.00 &  26.45 & 142.03 &  67.18 \\
294.1 & 241.78 &  57.08 & 126.47 &  94.88 &  7.8 & -1.3\hspace{5 pt} & 3.00$\times10^{-2}$ & 41.80$\pm$0.19 & 43.90$\pm$0.19 &  2.82 & 0.98 & 1.14 &  63.99 & 180.46 & 293.99 \\
294.1 & 181.55 & -26.23 & 106.55 &  95.99 & 20.6 &  2.3\hspace{5 pt} & 6.50$\times10^{-4}$ & 66.06$\pm$1.57 & 43.23$\pm$0.32 &  2.36 & 0.90 & 1.07 & 138.03 &  32.90 & 114.00 \\
294.2 & 207.78 &  34.43 & 117.53 &  91.85 & 32.7 & -0.7\hspace{5 pt} & 4.95$\times10^{-3}$ & 60.00$\pm$0.08 & 43.88$\pm$0.48 &  2.94 & 0.92 & 1.13 & 107.93 & 207.61 & 294.02 \\
343.3 & 250.03 &  64.42 & 114.32 & 102.70 & 24.5 &  2.6\hspace{5 pt} & 2.90$\times10^{-3}$ & 34.42$\pm$0.09 & 42.44$\pm$0.02 &  1.45 & 1.00 & 1.01 &  50.25 & 189.42 & 343.17 \\
 33.1 & 260.25 &  39.42 & 111.78 &  90.08 & 30.5 & -0.8\hspace{5 pt} & 1.79$\times10^{-2}$ & 41.71$\pm$0.23 & 42.37$\pm$0.31 &  1.14 & 0.91 & 1.03 &  63.98 & 218.31 &  32.94 \\
 50.5 & 295.15 &  36.58 & 108.11 &  95.05 & 15.2 &  2.3\hspace{5 pt} & 6.10$\times10^{-4}$ & 49.52$\pm$1.07 & 42.56$\pm$0.49 &  0.91 & 0.99 & 1.06 &  86.89 & 193.74 &  50.33 \\
134.7 & 344.12 &  50.05 & 107.51 &  85.89 & 61.0 & -0.8\hspace{5 pt} & 9.49$\times10^{-3}$ & 47.88$\pm$0.22 & 43.46$\pm$0.44 &  3.67 & 0.86 & 1.14 &  87.32 & 223.96 & 134.52 \\
136.6 &  42.29 &  59.85 & 133.10 & 107.51 &  8.3 &  0.8\hspace{5 pt} & 5.61$\times10^{-3}$ & 60.88$\pm$0.04 & 43.80$\pm$0.56 &  1.31 & 0.96 & 1.18 & 110.07 & 153.62 & 136.45 \\
144.2 &  63.58 &  60.14 & 117.26 &  95.14 & 10.8 &  0.8\hspace{5 pt} & 2.35$\times10^{-3}$ & 59.66$\pm$0.40 & 42.62$\pm$0.53 &  1.41 & 0.87 & 1.06 & 111.76 & 138.00 & 144.07 \\
\enddata
\tablecomments{$\lambda_{\rm S}$ is the solar longitude, $\alpha_{\rm R}$ and $\delta_{\rm R}$ are right ascension and declination of the radiant, $M_{\rm R}$ is the peak magnitude in $R$-band, $m$ is the mass assuming $\tau=0.7\%$, $\sigma$ is how many errors $v_{\rm h}$ is above the hyperbolic limit, and $q$, $e$, $i$, $\omega$, and $\Omega$ are the orbital elements.}
\end{deluxetable}

\begin{figure}
\plottwo{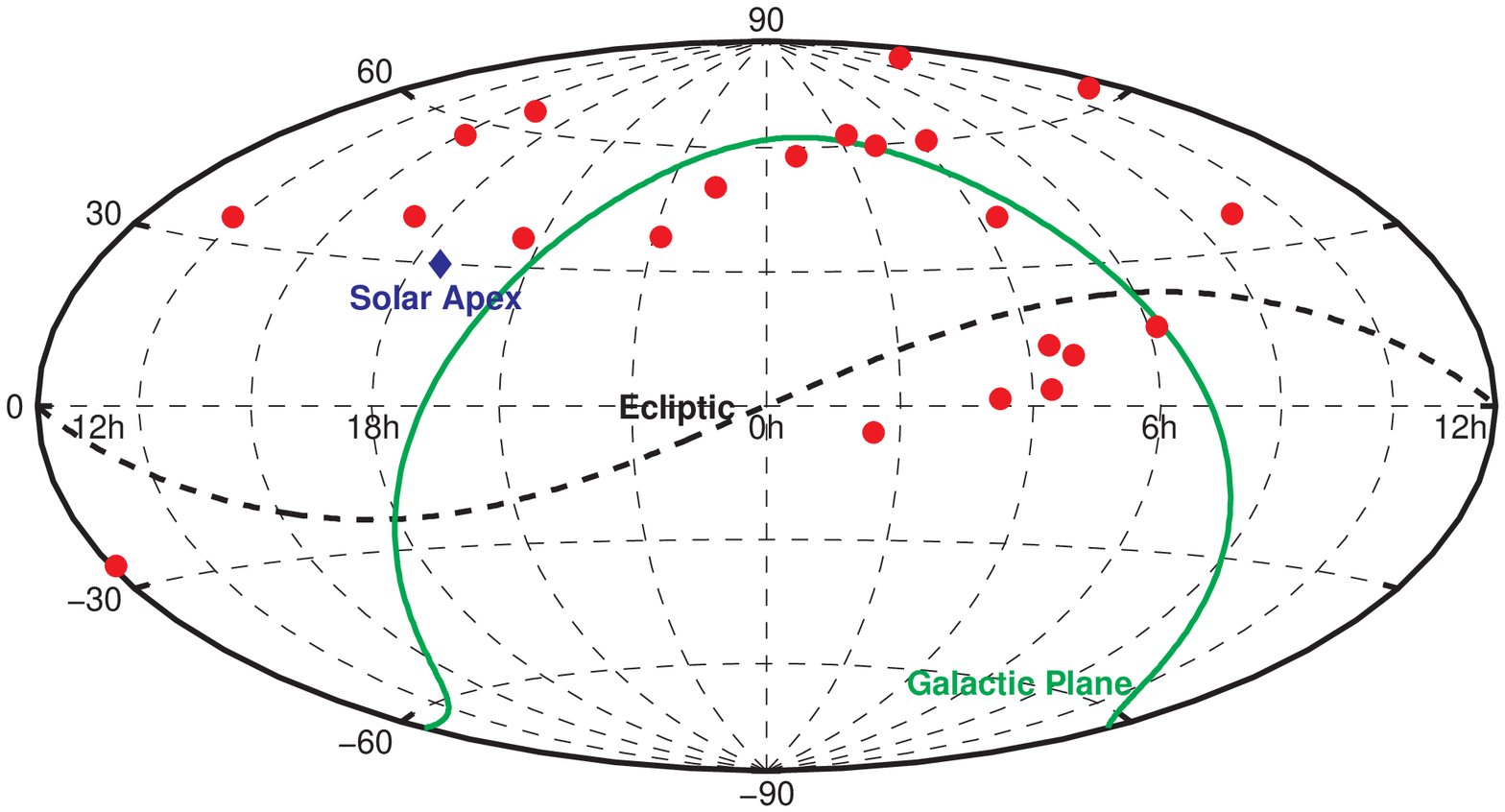}{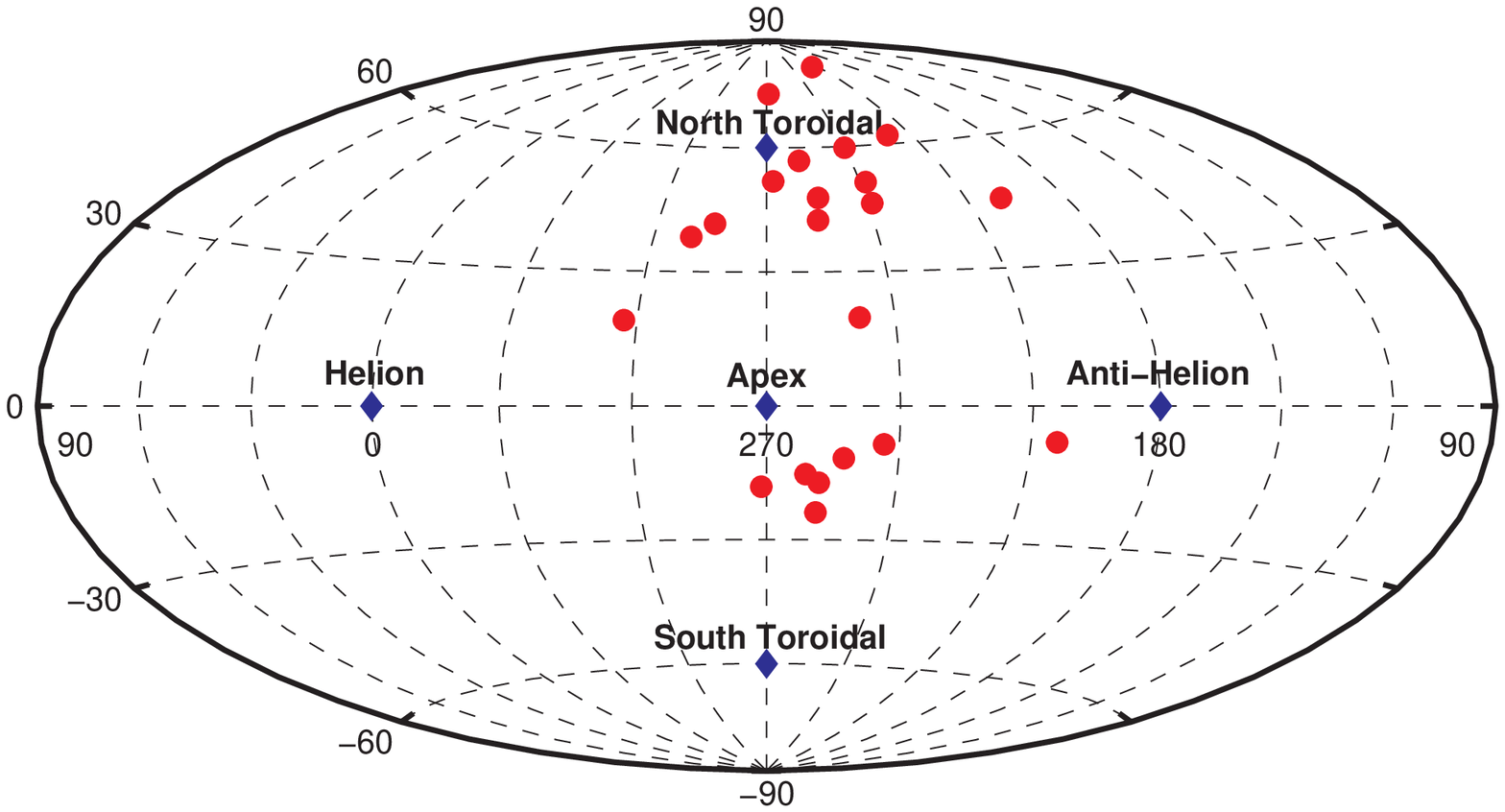}
\caption{Radiants for the 22 potential hyperbolic meteors. {\it Left:} Radiants in equatorial coordinates. {\it Right:} Sun centered ecliptic radiants in heliocentric coordinates. The origin is the apex of the Earth's motion and the Sun is at (0,0).\label{fig2}}
\end{figure}

We also analyzed the arrival directions in galactic coordinates for the 22 potential hyperbolic meteors. Genuine, large ISPs should have asymptotic arrival directions concentrated towards the solar apex and somewhat in the galactic plane in general \citep{mur2004}. We find no evidence for clustering of our orbital asymptote directions near the solar apex or the galactic plane (see Figure~\ref{fig2}\notetoeditor{Figure 2 should be full page width in print}, {\it left}), though the statistical data is sparse. The radiants in the galactic plane are the ones that could be associated with meteor showers. Figure~\ref{fig2} ({\it right}) shows that most radiants are from the toroidal sporadic source direction \citep{j&b1993}, which suggests an interplanetary origin. To calculate an extreme upper limit for the ISP flux at Earth we therefore used $N_{\rm ISP}=1$, $t=300$\,h, and $A_{\rm col}=16$\,km$^2$ and get $f_{\rm ISP}<2\times 10^{-4}$\,km$^{-2}$\,h$^{-1}$.

The 22 potential hyperbolic meteors are all close to the hyperbolic limit at Earth's orbit. Observational uncertainties are a distinct probable cause of the apparent unbound orbits (see Section~\ref{s_ana}). Another possibility is that the meteoroids recently suffered a close encounter with a planet, which increased their orbital energies. This gravitational slingshot effect has been shown to be capable of accelerating meteoroids near the heliocentric escape velocity onto hyperbolic orbits, and that such meteoroids may intersect the Earth before leaving the Solar System \citep{wie2011}.

To examine this possibility, the orbits of the 22 candidates were integrated backwards for fifty years. Along with the nominal meteoroid orbit, a suite of 100 clones with initial positions and velocities chosen from within the one-sigma uncertainty bounds were followed. From this we can address the question of whether or not any of these reliably measured orbits are consistent with having been produced by a planetary encounter.

The integrations are performed with the RADAU algorithm \citep{eve1985} in two stages.  A five minute step is used for the 24 hours immediately preceding the meteoroids arrival at Earth: such a small time step is necessary to adequately account for the deflection of the orbit produced by Earth's gravity. After that point, the time step is changed to one day and the integration is extended 50 years backwards. The output is scrutinized for close approaches with the planets, measured in units of Hill radii.

The simulations include the planets Mercury through Neptune. Earth and Moon are treated as separate bodies. When the check for close approaches is made, the Moon is the most common result. However, this is a simple result of the Moon's position approximately six Hill radii from Earth: any meteor that strikes Earth must pass at least this distance from the Moon. Close approaches to the Moon at this distance have a negligible effect on the meteoroids orbit, and we concentrate our attention on those meteoroids that may have suffered close encounters with one of the more massive bodies of the Solar System.

For the 22 hyperbolic meteors simulated, five had one or more clones that had close encounters with a planet just prior to their arrival at Earth. None of the simulated encounters produced marked changes in their orbits. However, this is expected given the small number of clones simulated. What this really indicates is that these five meteors' trajectories are consistent within the measurement uncertainties of having encountered a planet prior to arrival at Earth. This does not mean that they were certainly accelerated to hyperbolic speeds by such an encounter; we do not have enough information to determine this with certainty. It does imply, however, that the majority of our nominally hyperbolic events could not have been more than very slightly perturbed due to a recent close planetary encounter.

\section{Discussion and Conclusions}

Our survey conducted between 2009 June 4 and 2010 August 17 contains 1739 meteors observed from two sites. A hyperbolic orbit was measured for 115 of these events. Only 22 of those have a reliable solution, i.e., enough measured data points, a convergence angle $Q>5\degr$, and $<$2$\%$ average velocity difference from the trajectory solutions of the two stations. The heliocentric speed is $3\sigma$ above the hyperbolic limit for only two meteors with a reliable solution, though in both cases the actual speed above the hyperbolic limit was $<$\,2.2\,km\,s$^{-1}$.

For true ISPs we expect a heliocentric velocity $v_{\rm h,ISP}\gtrsim46$\,km\,s$^{-1}$, based on typical stellar speeds of $20$\,km\,s$^{-1}$ with respect to the Solar System. None of the 22 potential hyperbolic meteors had $v_{\rm h}>45$\,km\,s$^{-1}$. But the difference between the expected $v_{\rm h,ISP}$ and the measured $v_{\rm h}$ is within $1\sigma$ in one case. The incoming directions, however, are not clearly from the solar apex or the galactic plane as we expect for ISPs, but rather clustered along the ecliptic as expected for interplanetary meteoroids. Hence, we find no clear indication of an interstellar signature within the processed data, though the number statistics for possible ISPs is small.

It has to be noted that for some encounter geometries interstellar meteors may have large velocities. Such meteors would appear on a few frames only with our system and may be rejected because of too few measured points. Furthermore, the system configuration is optimized to detect meteors at an altitude of 100\,km, with meteors being detected up to 125\,km altitude. ISPs radiating at much higher altitudes would not be detected from both sites. We need, however, good quality control in order to identify true interstellar meteoroids and expect only a minority of the true ISPs encountering Earth to have significantly higher speeds than interplanetary meteoroids (see \citet{b&n2002} for a discussion).

Another issue is that both, the galactic center and the debris disk star $\beta$ Pic, which has been identified as possible source of hyperbolic meteors \citep{bagg2000}, are only visible from the southern hemisphere. One would expect more ISPs coming from the direction of the galactic center as the density of stars is higher. However, due to the large distance of the solar system to the galactic center the contribution of ISPs from stars in the solar neighborhood is probably dominant. It is also not obvious why $\beta$ Pic should be the only source of ISPs. Other debris disk stars that are comparable to $\beta$ Pic, e.g., $\alpha$ Lyr \citep{back1997}, can be seen from the northern hemisphere. We therefore assume that we can observe a typical ISP flux from the location of our sites.  

We obtained an upper limit on the influx at Earth of ISPs with masses $m>2\times10^{-7}$\,kg of $f_{\rm ISP}<2\times10^{-4}$\,km$^{-2}$\,h$^{-1}$. This is clearly below of what is expected if we extend the size distribution model from \citet{mrn1977} to larger particles or simply apply the collisional cascade model from \citet{doh1969}, anchored from the reliable ISP fluxes measured by Ulysses and Galileo. Consequently, the slope for larger masses has to be steeper than in their model. \citet{mrn1977}, however, only looked at particles smaller than $\sim$10$^{-15}$\,kg. \citet{wya2007} indicate that particles smaller than a certain diameter $D_{\rm bl}$ are blown out by radiation pressure as soon as they are created. We can therefore expect that ISPs larger than $D_{\rm bl}$ are underrepresented compared to the collisional cascade model. The blowout diameter in $\mu$m is defined as \citep{wya2007}
\begin{equation}
D_{\rm bl} = 0.8\frac{L_{*}}{M_{*}}\frac{2700}{\rho},
\end{equation}
where $L_{*}$ and $M_{*}$ are the stellar luminosity and mass, respectively, in solar units and $\rho$ is the density of the particle in kg\,m$^{-3}$. To estimate an upper limit for the mass of ISPs that follow the collisional cascade model we assume $\rho=1000$\,kg\,m$^{-3}$. We get $D_{\rm bl}=2.16\,L_{*}/M_{*}$ and an estimated blowout mass of
\begin{equation}
M_{\rm bl} \approx 10^{-14}\left(\frac{L_{*}}{M_{*}}\right)^3,
\end{equation}
where $M_{\rm bl}$ is in kg. The blowout mass for the Sun is $M_{\rm bl}\approx10^{-14}$\,kg, for $\beta$ Pic $M_{\rm bl}\approx10^{-12}$\,kg, and for Sirius, the most massive star in the solar neighborhood, $M_{\rm bl}\approx2\times10^{-11}$\,kg. The exponent $q$ in the power-law fit must therefore be larger than $q=0.83$ for $m\gtrsim10^{-11}$\,kg. This roughly matches with the break in the fit from \citet{haj2002}.

If we combine our result with other ISP flux estimates and examine other possible power-law fits (see Figure~\ref{fig1}), $q=1.1$ from \citet{lan2000} seems to be the most promising representation of the true flux for masses $m\gtrsim10^{-7}$\,kg, as our estimated flux is an upper limit. The fit from both, \citet{lan1998} ($q=0.9$) and \citet{haj2002}, would predict a higher flux than we observe. Even so, we cannot completely rule out the fit from \citet{haj2002} as a possible solution if we consider uncertainties in their result. With the current CAMO system, we would need to observe at least 800 hours in order to confirm or reject $q=1.1$ from \citet{lan2000}. However, it is possible to combine several narrower FOV electro-optical instruments to detect meteoroids over 10 times less massive than the presented sample and at the same time observing a similar collecting area using a system similar to the one described in \citet{kik2009}. A suite of such cameras at two sites could be more than an order of magnitude more efficient in searching for ISPs than our current system as well as more precise. This offers the most promising near-term prospect for unambiguous ISP detection at Earth.  

\acknowledgments

We thank the Canadian Foundation for Innovation, the Natural Sciences and Engineering Research Council of Canada, and the NASA Meteoroid Environment Office for funding support and J.\,Borovi{\v c}ka and Z.\,Ceplecha for use of their processing software. We acknowledge the technical assistance of Z. Krzeminski in maintaining the CAMO system. We thank the referee, J.\,Borovi{\v c}ka, who provided constructive comments that helped to improve the paper.

\end{document}